# Optimal Constrained Resource Allocation Strategies under Low Risk Circumstances

Mugurel Ionut Andreica *, Madalina Ecaterina Andreica **, Costel Visan ***

\* Politehnica University of Bucharest, Bucharest, Romania, email: mugurel.andreica@cs.pub.ro
\*\* The Bucharest Academy of Economic Studies, Bucharest, Romania, email: madalina.andreica@gmail.com
\*\*\* The Bucharest Academy of Economic Studies, Bucharest, Romania, email: costel.visan@yahoo.com

## ABSTRACT

In this paper we consider multiple constrained resource allocation problems, where the constraints can be specified by formulating activity dependency restrictions or by using game-theoretic models. All the problems are focused on generic resources, with a few exceptions which consider financial resources in particular. The problems consider low-risk circumstances and the values of the uncertain variables which are used by the algorithms are the expected values of the variables. For each of the considered problems we propose novel algorithmic solutions for computing optimal resource allocation strategies. The presented solutions are optimal or near-optimal from the perspective of their time complexity. The considered problems have applications in a broad range of domains, like workflow scheduling in industry (e.g. in the mining and metallurgical industry) or the financial sector, motion planning, facility location and data transfer or job scheduling and resource management in Grids, clouds or other distributed systems.

## KEYWORDS
Resource Allocation, Workflow Scheduling, Directed Tree, Game-Theoretic Models, Geometric Constraints.

## 1. INTRODUCTION

In this paper we address several constrained resource allocation problems, in which the problem parameters are considered to present low fluctuation risks. For instance, each parameter $P$ may have an associated probability distribution (e.g. Gaussian, discrete), which models the probability associated to each value $v$ that the parameter $P$ may take. In the case of a discrete probability distribution, $P$ may take $NV(P)$ different values $v_1, …, v_{NV(P)}$, each value $v_i$ having an associated occurrence probability $prob_i$ (with $prob_1+…+prob_{NV(P)}=1$). Because of the low fluctuation risk, we will compute the expected value $E(P)$ of the parameter $P$ (in the case of a discrete distribution, $E(P)=v_1 \cdot prob_1+…+v_{NV(P)} \cdot prob_{NV(P)}$) and we will only consider this value in the following problems. Using the expected value of a parameter $P$ instead of the probability distribution simplifies matters enough such that for the considered problems we were able to develop efficient algorithmic solutions. In the considered problems we will implicitly assume that the given values of the parameters are expected values. We should keep in mind, though, that each problem can be also modeled by assigning probability distributions to its parameters, instead of fixed expected values. However, handling probability distributions is much more difficult than handling expected values and, thus, in this paper we only consider expected values.

In Section 2 we consider a cost optimization problem which may occur in several situations. We have a business workflow [10] in which the dependencies between activities are specified as a directed tree. Every activity produces its own output, which needs to be stored, in order to be used by the activity which depends on it. We have two different types of storage with different storage costs and we are interested in finding a data storage strategy which minimizes the total costs. In Section 3 we consider a debt management problem in which the customer needs to repay his debts to several banks by distributing his assets to the banks; however, the assets do not have a fixed value and each bank may perceive each asset as having a potentially different value. In Section 4 we use game-theoretic models for modeling several resource allocation problems and in Section 5 we consider geometric optimization problems, in the context of modeling resources as points in a multidimensional attribute space. In Section 6 we present related work and in Section 7 we conclude and discuss future work.

## 2. MINIMIZING DATA STORAGE COSTS IN DIRECTED TREE WORKFLOW SCHEDULING

We consider a workflow which is structured as a directed rooted tree (with root $r$) consisting of $N$ activities (vertices). We denote by $ns(i)$ the number of sons of a vertex $i$, by $s(i,j)$ ($1 \leq j \leq ns(i)$) the $j^{th}$ son of vertex $i$ (in some arbitrary order) and by $parent(i)$ the parent of vertex $i$ ($parent(r)=undefined$). We also denote by $T(i)$ the subtree rooted at vertex $i$. The activity (vertex) $i$ cannot be executed before the activities (vertices) $s(i,1), …, s(i,ns(i))$ are executed (i.e. vertex $i$'s sons), because of data dependency issues. Every vertex has at most $K$ sons, where $K$ is a small value.

We consider that the output produced by every activity has the same size (*1* unit). There are two types of storage systems. The first one ($S_1$) can store at most $D$ data units at a time, at *zero* cost (e.g. the storage system of the workflow manager). The second storage system ($S_2$) can store any amount of data, at different costs. In order to store the output of activity $i$ in $S_2$, the cost which needs to be paid is $C(i)$ (which may be positive or negative).

After an activity $i$ produces its output, the output must be stored in $S_1$ or $S_2$ until it is required by *parent(i)* (for $i=r$, the output does not need to be stored anywhere). We want to compute a serial schedule (an execution order for the activities) and decide for each activity output where to store it, such that the total costs paid in order to store data in $S_2$ are minimized.

An efficient solution (for the case when all the costs $C(i)$ are equal and positive), based on the register allocation algorithm presented in [14], was given as a solution to a problem proposed at the Baltic Olympiad in Informatics 2003. For each vertex $i$, we compute $DN(i)$=the minimum amount of data units required in $S_1$ in order to execute all the activities in vertex $i$'s subtree, such that no output has to be stored in $S_2$. For a leaf vertex $l$, we have $DN(l)=0$. For a non-leaf vertex $i$, we sort its sons, such that $DN(s(i,1)) \geq DN(s(i,2)) \geq ... \geq DN(s(i,ns(i)))$. $DN(i)=max\{ns(i), max\{DN(s(i,j))+j-1|1 \leq j \leq ns(i)\}\}$. This step takes $O(N \cdot log(K))$ time overall. If $D \geq DN(1)$, the total cost is $0$. Otherwise, we define a function $C_{min}(i,Q)$ which computes the minimum cost required for executing activity $i$, if only $Q$ data units are available in $S_1$ (we do not consider the storage cost for the output of activity $i$); if $Q \geq DN(i)$, then $C_{min}(i,Q)=0$; if $Q<0$, then $C_{min}(i,Q)=+\infty$. We want to compute $C_{min}(r,D)$. In order to compute $C_{min}(i,Q)$, we need to find a subset $ss2(i)$ and an ordering $sp(i,1), ..., sp(i,ns(i)-|ss2(i)|)$ of the sons $s(i,j) \in ss1(i)=\{s(i,1), ..., s(i,ns(i))\} \setminus ss2(i)$, such that: (1) all the activities of the sons $s(i,j) \in ss2(i)$ will be executed first, and their results will be stored in $S_2$; every such activity will have $Q$ data units available in $S_1$; (2) then, we execute the activities of the sons $s(i,j') \in ss1(i)$, in the order $sp(i,1), ..., sp(i,|ss1(i)|)$ and store their results in $S_1$; son $sp(i,j)$ will have $Q-j+1$ data units available in $S_1$. The total cost for $C_{min}(i,Q)$ for the chosen parameters is the sum of the values $(C(s(i,j))+C_{min}(s(i,j),Q))$ (with $s(i,j) \in ss2(i)$), plus the sum of the values $C_{min}(sp(i,j), min\{DN(sp(i,j)), Q-j+1\})$ ($1 \leq j \leq |ss1(i)|$). If we store all the computed values $C_{min}(*,*)$ and we consider every possible subset $ss2(i)$ and every permutation of the sons in $ss1(i)$, we obtain a time complexity of $O(N \cdot D \cdot K! \cdot 2^K)$. However, we can do better by using the following observations. For a son $sp(i,j)$ ($2 \leq j \leq |ss1(i)|$), if $Q-j+1<DN(sp(i,j))$, then we can move $sp(i,j)$ from $ss1(i)$ to $ss2(i)$, without increasing the total cost. Thus, for every son $sp(i,j)$ ($2 \leq j \leq |ss1(i)|$) from $ss1(i)$, we must have $DN(sp(i,j)) \leq Q-j+1$. With this observation, we notice that we can always sort the sons $sp(i,j)$ from $ss1(i)$ such that $DN(sp(i,1)) \geq DN(sp(i,2)) \geq ... \geq DN(sp(i,|ss1(i)|))$. With these observations, we only need to find the optimal set $ss2(i)$. Once this set is found, the ordering $sp(i,1), ..., sp(i,|ss1(i)|)$ of the vertices in $ss1(i)$ is fixed, we have $DN(sp(i,j)) \leq Q-j+1$ (for $2 \leq j \leq |ss1(i)|$), and the cost $C_{min}(i,Q)$ is equal to the sum of the values $(C(s(i,j))+C_{min}(s(i,j),Q))$ (with $s(i,j) \in ss2(i)$), plus $C_{min}(sp(i,1),Q)$, plus the sum of the values $C_{min}(sp(i,j), DN(sp(i,j)))$ ($2 \leq j \leq |ss1(i)|$). In order to compute the optimal value $C_{min}(i,Q)$, we consider the sons of vertex $i$ sorted such that $DN(s(i,1)) \geq ... \geq DN(s(i,ns(i)))$ and we run a dynamic programming algorithm. We compute $C_{aux,min}(i,Q,j,p)$=the minimum total cost if we considered the first $j$ sons so far and $p$ of them ($0 \leq p \leq j$) were inserted into $ss1(i)$. We have $C_{aux,min}(i,Q,0,0)=0$. For $1 \leq j \leq ns(i)$, we have the following equations: (1) $C_{aux,min}(i,Q,j,0)=C_{aux,min}(i,Q,j-1,0)+C(s(i,j))+C_{min}(s(i,j),Q)$ ; (2) $C_{aux,min}(i,Q,j,j)=C_{aux,min}(i,Q,j-1,j-1)+C_{min}(s(i,j), min\{Q-j+1, DN(s(i,j))\})$ (if $j=1$ or $Q-j+1 \geq DN(s(i,j))$), or $+\infty$ (if $j>1$ and $Q-j+1<DN(s(i,j))$); (3) for $1 \leq p \leq j-1$, we have: $C_{aux,min}(i,Q,j,p)=min\{C_{aux,min}(i,Q,j-1,p)+C(s(i,j))+C_{min}(s(i,j),Q), C_{aux,min}(i,Q,j-1,p-1)+C_{min}(s(i,j), min\{Q-p+1, DN(s(i,j))\}$ (if $p=1$ or $Q-p+1 \geq DN(s(i,j))$) or $+\infty$ (if $p>1$ and $Q-p+1<DN(s(i,j)))\}$. We have $C_{min}(i,Q) = min\{C_{aux,min}(i,Q,ns(i),p)|0 \leq p \leq ns(i)\}$. The dynamic programming stage takes $O(ns(i)^2)$ time for a vertex $i$. Apparently, we need to compute $C_{min}(i,Q)$ for every vertex $i$ and every value $Q$ ($0 \leq Q \leq D$), obtaining an $O(N \cdot D \cdot K^2)$ time complexity. However, we must notice that, when computing $C_{min}(i,Q)$, we only need to know the values $C_{min}(s(i,j),Q)$ and the values $C_{min}(s(i,j), DN(s(i,j)))$ (the second set of values are always $0$) ($1 \leq j \leq ns(i)$). Since we are only interested in $C_{min}(r,D)$, we only need to compute the values $C_{min}(i,D)$ (besides the already known values $C_{min}(i,DN(i))$), obtaining an $O(N \cdot K^2)$ time complexity.

For the general case (with not necessarily equal or positive $C(i)$ values), we will traverse the tree vertices bottom-up. For each vertex $i$ we will compute $C_{min}(i,j)$=the minimum cost required for executing activity $i$, if only $j$ data units are available in $S_1$ (ignoring the cost of storing activity $i$'s output anywhere). For a leaf node $l$, $C_{min}(l,j)=0$ (for $0 \leq j \leq D$). For a non-leaf vertex $i$, we will consider a variable $x(i)$ ranging from $0$ to $ns(i)$. $x(i)$ represents the number of sons whose output data will be stored in $S_1$. The other $ns(i)-x(i)$ sons (forming the set $S(i, x(i))$) will have their output stored in $S_2$. Each of the sons in $S(i,x(i))$ will have all the $j$ data units from $S_1$ available at the start of the execution of the activities in their subtrees (after executing one of these sons, no extra data is stored in $S_1$). Let $o(1), ..., o(x(i))$ be the order of the $x(i)$ sons whose output is stored in $S_1$. Son $o(1)$ has $j$ data units available in the beginning. Son $o(2)$ will have $j-1$ data units, ..., son $o(p)$ will have $(j-p+1)$ data units available ($1 \leq p \leq x(i)$). Thus, for a fixed value $x(i)$, we need to compute the set $S(i, x(i))$ and the order $o(1), ..., o(x(i))$, such that the total cost is minimum. A son $y \in S(i,x(i))$ contributes to the total cost with a value equal to $C_{min}(y,j)+C(y)$. Son $o(p)$ ($1 \leq p \leq x(i)$) contributes to the total cost with a value equal to $C_{min}(o(p), j-p+1)$. A first solution consists of considering all the $C(ns(i), ns(i)-x(i))$ possibilities of choosing the sons in $S(i,x(i))$

($C(a,b)$=combinations of *a* elements taken *b* at a time) and all the *(x(i))!* ordering possibilities for the sons outside of *S(i,x(i))*. The complexity of this solution is $TC(K)=O(K!+C(K,1)\cdot(K-1)!+C(K,2)\cdot(K-2)! + ... + C(K,q)\cdot(K-q)!+ ... + C(K,K)\cdot 0!)$ for a vertex *i* and *j* available data units in $S_1$. Thus, the overall time complexity is $O(N\cdot D\cdot TC(K))$. The minimum total cost is $C_{min}(r, D)$.

A more efficient solution is the following. For every value of *x(i)* we will construct a bipartite graph. The left side of the graph contains the sons of vertex *i* and the right side contains *x(i)+1* vertices. The first *x(i)* vertices *p* on the right side have the meaning that *(j-p+1)* data units will be available ($1\leq p\leq x(i)$). The last vertex *(p=x(i)+1)* means that the son will be placed into *S(i,x(i))*. We have directed edges between every son *y* on the left side and every vertex *p* on the right side. If $1\leq p\leq x(i)$, the cost of such an edge will be $C_{min}(y, j-p+1)$ and its capacity will be *1*; for *p=x(i)+1*, the cost of the edge will be $C_{min}(y, j)+C(y)$ and its capacity will also be *1*. We will insert an extra node *src*, with directed edges from it to every vertex on the left side of the bipartite graph (of zero cost and unit capacity), and an extra node *dst*, with zero cost directed edges from every vertex on the right side of the bipartite graph and *dst*. The edges *(p,dst)* ($1\leq p\leq x(i)$) have unit capacity. The edge *(x(i)+1,dst)* has a capacity equal to *(ns(i)-x(i))*. We will compute a minimum cost maximum flow in this graph. Let *CC(i, j, x(i))* be the cost of such a flow. $C_{min}(i,j)=min\{CC(i,j,x(i))|0\leq x(i)\leq ns(i)\}$. Let's analyze now the time complexity of the obtained algorithm. Every graph where a minimum cost maximum flow computation takes place (for every tuple *(i, j, x(i))*) has *O(K)* vertices. A well-known method for computing such a flow consists of performing *O(K)* iterations and, at each iteration, we will compute a minimum-cost path from *src* to *dst*, in the residual graph, where some edges may have negative costs. The Bellman-Ford (or Bellman-Ford-Moore) algorithm can compute such a path in $O(K^3)$ time, obtaining an $O(K^4)$ time complexity for the flow computation. However, because of the particular nature of the graph, we can also use Dijkstra's algorithm (even if edges with negative costs exist in the residual graph, the algorithm will work correctly because we can perform a transformation which makes the cost of every edge non-negative). Thus, a minimum cost maximum flow can be computed in $O(K^3)$ time ($O(K^2)$ per iteration). Since we have $O(N\cdot D\cdot K)$ tuples *(i, j, x(i))*, the overall time complexity may seem to be $O(N\cdot D\cdot K^4)$. However, for every vertex *i* with *ns(i)>1* sons, there will be at least *ns(i)-1* leaves in *T(i)*. Thus, there are at most *O(N/K)* vertices with *O(K)* sons. With this observation, the time complexity is, in fact, $O(N/K\cdot D\cdot K^4)=O(N\cdot D\cdot K^3)$.

## 3. DEBT MANAGEMENT WITH ASSETS OF DIFFERENTLY PERCEIVED VALUES

We consider the following problem. A customer owes *P(i)* euro to each of the *d* banks from which he obtained a credit ($1\leq i\leq d$). In order to repay the debt, the customer will need to give up on some of his assets. He has $2^d$ types of assets (numbered from *0* to $2^d-1$). Each bank perceives every asset as having a value of either *1* or *2* euro. If an asset is of type *T*, then we will consider the binary representation of *T*: <u>*b(1)b(2)...b(d)*</u>. If *b(i)=0*, then bank *i* perceives this asset as having a *1* euro value; if *b(i)=1*, bank *i* perceives the asset as having a *2* euro value. The customer has *C(T)* assets of type *T*. All of the assets must be distributed to the banks, in such a way that the total value of the assets perceived by each bank *i* is at least *P(i)*.

The main idea of the solution is to maximize the *utility* of the asset distribution. An asset *x* is useful if it is distributed to a bank *i* which perceives it as having a value of *2* euro and the total value of the assets distributed to that bank, plus the value of the asset *x* is at most *P(i)*. We will solve this problem in two stages. During the first stage we will construct a bipartite graph which has $2^d$ vertices on the left side and *d* vertices on the right side. We add a directed edge between every vertex *u* ($0\leq u\leq 2^d-1$) on the left side and every vertex *v* ($1\leq v\leq d$) on the right side, only if bank *v* perceives the asset of type *u* as having a value of *2* euro. The capacity of the edge will be $+\infty$. We then add a source *S* and a sink *Q*. We add directed edges from the source *S* to every vertex *u* on the left side (the capacity of such an edge will be *C(u)*). Then we add directed edges from every vertex *v* on the right side and the sink *Q* (each such edge will have capacity *(P(v) div 2)*). We will now compute a maximum flow from *S* to *Q* in this network. For every edge *(u,v)* (*u* on the left side and *v* on the right side) with a flow *f(u,v)* on it, we will distribute *f(u,v)* assets of type *u* (initially) to the bank *v* and we will decrease *P(v)* by *2·f(u,v)* and *C(u)* by *f(u,v)*. In the second stage, we traverse the banks in increasing order (*i=1,...,d*) and while *P(i)>0*, we choose an asset of any type *T* with *C(T)>0* and give it to bank *i*; after this, we decrease *P(i)* by *val(i,T)* (the value perceived by bank *i* for the asset of type *T*) and *C(T)* by *1*. In the final stage, when all the banks have received assets which fully cover the debts, all the remaining non-distributed assets are distributed to any of the banks.

Solutions which do not necessarily make use of the maximum flow computation can be obtained for *d=1,2,3*. The case *d=1* is trivial (bank *1* receives all the assets). For *d=2*, we give as many assets of type <u>*10*</u> to bank *1* and as many assets of type <u>*01*</u> to bank *2* as possible (without exceeding the limits *P(1)* and *P(2)*). Then, we consider the banks *i* in any order (e.g. *i=1,2*) and, as long as we still have assets of type <u>*11*</u> and the total perceived sum of bank *i* is at most *P(i)-2*, we give an asset of type <u>*11*</u> to bank *i* and increment the total perceived sum of bank *i* by *2*. In the end, we consider the banks *i* and, while the perceived sum of the received assets is smaller than *P(i)*, we give to bank *i* an asset of type <u>00</u> or <u>11</u> (whichever is still available) and increment the total perceived sum of bank *i* by the corresponding value (*1* or *2*).

For *d=3*, we will perform several stages and we will maintain the values *S(i)*=the total value of the assets distributed to bank *i* (as perceived by bank *i*). Initially, *S(i)=0* ($1 \leq i \leq 3$). *C(T)* will be the number of assets of type *T* the customer still possesses (these numbers will be decremented during the course of the algorithm). In the first stage we will distribute assets of the type *t(1,1)=<u>100</u>* (to bank *1*), *t(1,2)=<u>010</u>* (to bank *2*) and *t(1,3)=<u>001</u>* (to bank *3*). If *2·C(t(1,i))≤P(i)*, then we give all the type *t(1,i)* assets to bank *i*: we set *S(i)=2·C(t(1,i))* and then *C(t(1,i))=0*; if *2·C(t(1,i))>P(i)*, we will compute *q=P(i) div 2* (integral division) and we will set: *S(i)=2·q* and *C(t(1,i))=C(t(1,i))-q*.

The second step is the most important. We will distribute assets of the types <u>011</u>, <u>101</u> and <u>110</u>, in such a way that their total *utility* is maximum. An asset is useful if it is distributed to a bank *i* which perceives as having value *2* and for which *S(i)* does not exceed *P(i)* after receiving the asset. We will consider, one at a time, every value of *x* from *0* to *min{C(<u>110</u>), (P(1)-S(1)) div 2}* and we will assume that bank *1* receives *x* assets of the type <u>110</u>. Let's consider the values *S'(x,i)* and *C'(x,T)*, having the same meaning as *S* and *C*, but for the „virtual" case in which bank *1* receives *x* assets of type <u>110</u>. Initially, we have *S'(x,i)=S(i)* ($1 \leq i \leq 3$) and *C(x,T)=C(T)* (*T=<u>011</u>, <u>101</u>* or <u>110</u>). We will increment *S'(x,i)* by *2·x* and we will decrement *C'(x, <u>110</u>)* by *x*. All the other type <u>110</u> assets are useful only for bank *2*. We will consider the following procedure *GiveMax(x,T,i)*: if *C'(x,T)·2≤P(i)-S'(x,i)*, we increment *S'(x,i)* by *C'(x,T)·2* and we set *C'(x,T)* to *0*; otherwise, we increment *S'(x,i)* by *2·q*, where *q=(P(i)-S'(x,i)) div 2*, and we decrement *C'(x,T)* by *q*. We will call *GiveMax(x, <u>110</u>, 2)*. Then, we will consider the type <u>011</u> assets. We will give these assets to bank *2*, as long as they are useful, by calling *GiveMax(x, <u>011</u>, 2)*. The rest of type <u>011</u> assets are only useful to bank *3*; thus, we will call *GiveMax(x, <u>011</u>, 3)*. We now get to the type <u>101</u> assets. We distribute as many of these to bank *3* as possible, with the condition that they are useful, by calling *GiveMax(x, <u>101</u>, 3)*. Then we call *GiveMax(x, <u>101</u>, 1)* (in order to give the rest of the type <u>101</u> assets to bank *1*, as long as they are useful). After all these computations, we define *U(x)*=the sum of the values *(S'(x,i)-S(i))* ($1 \leq i \leq 3$) (i.e. the total utility for the case when bank *1* receives *x* assets of type <u>110</u>). After considering every value of *x*, we will choose that value *xmax*, for which *U(xmax)* is maximum. We will perform all the actions corresponding to *xmax* and, afterwards, we will set *S(i)=S'(xmax,i)* ($1 \leq i \leq 3$) and *C(T)=C'(xmax,T)* (*T=<u>011</u>, <u>101</u>, <u>110</u>*).

In the third stage we will distribute the assets of type <u>111</u> to the *3* banks, as needed. We consider every bank *i* ($1 \leq i \leq 3$) and if *S(i)<P(i)* then: *(1)* if *S(i)+2·C(T)≤P(i)*, then we set: *S(i)=S(i)+2·C(T)* and then *C(T)=0*; *(2)* otherwise, if *S(i)+2·C(T)>P(i)*, then we compute *q=(P(i)-S(i)) div 2* and we set: *S(i)=S(i)+2·q*, and then *C(T)=C(T)-q*.

Within stage *4*, we will consider, one at a time, every bank *i* ($1 \leq i \leq 3$) and if *S(i)<P(i)*, we will consider each of the types *j* ($0 \leq j \leq 7$). Let *val(i,j)* be the value perceived by bank *i* for an asset of type *j*; *(1)* if *S(i)+C(j)·val(i,j)≤P(i)*, then we set: *S(i)=S(i)+C(j)·val(i,j)*, and then *C(j)=0*; *(2)* if, however, *S(i)+C(j)·val(i,j)>P(i)*, then we compute *q=(P(i)-S(i)) div val(i,j)*, and then we set *S(i)=S(i)+q·val(i,j)* and, after this, *C(j)=C(j)-q*; if, after these changes, we have *S(i)<P(i)* and *C(j)>0*, then we increment *S(i)* by *val(i,j)* and we decrement *C(j)* by *1* (we give an extra asset of type *j* to bank *i*). During the last stage of the algorithm we will distribute the remaining assets to any of the banks.

The case where the customer has *Q* assets overall and the banks may perceive the value of any asset *a* as being arbitrary (e.g. we have *val(i,a)*=the value of asset *a*, perceived by bank *i*) is equivalent to a multidimensional knapsack problem. We have two approaches. First, we can compute *OK(a, w(1), …, w(d))=true* or *false*, if we can reach a state in which assets of total value *w(j)* were distributed to bank *j* ($1 \leq j \leq d$), considering only the first *a* assets. We have *OK(0, 0, …, 0)=true* and *OK(0, w(1), …, w(d))=false* if we have at least one value *w(j)>0*. For $1 \leq a \leq Q$ we have *OK(a, w(1), …, w(d))=OR{OK(a-1, w(1), …, w(d)), OR{OK(a-1, w(1), …, w(j-1), w(j)-val(j,a), w(j+1), …, w(d)|$1 \leq j \leq d$, w(j)≥val(j,a)}}*. *OR(S)* represents the logical *OR* between an auxiliary boolean value equal to *false* and all the boolean values of the set *S*. If there is at least one value *OK(Q, w(1), …, w(d))=true* with *w(j)≥P(j)* (for every $1 \leq j \leq d$), then the assets can be distributed to the banks such that the debt is repaid. A second approach consists of computing $Val_{max}(a, w(1), …, w(d-1))$=the maximum total value of the assets distributed to bank *d*, if the total value of the assets distributed to bank *j* is *w(j)* ($1 \leq j \leq d-1$) and we considered only the first *a* assets. We have $Val_{max}(0, 0, …, 0)=0$ and $Val_{max}(0, w(1), …, w(d-1))=-\infty$, if we have at least one value *w(j)>0* ($1 \leq j \leq d-1$). For $1 \leq a \leq Q$ we have: $Val_{max}(a, w(1), …, w(d-1))=max\{Val_{max}(a-1, w(1), …, w(d-1))+max\{val(d,a), 0\}, max\{Val_{max}(a-1, w(1), …, w(j-1), w(j)-val(j,a), w(j+1), …, w(d-1)|1 \leq j \leq d-1, w(j)≥val(j,a)\}\}$. If there is at least one value $Val_{max}(Q, w(1), …, w(d-1))≥P(d)$ (with *w(j)≥P(j)* for every $1 \leq j \leq d-1$), then we have a solution. In both cases, the actual distribution of assets to the *d* banks can be computed by tracing back the way the *OK(\*, …, \*)* or the $Val_{max}(*, …, *)$ values were computed. The time complexities of these algorithms are pseudo-polynomial, if all the *P(\*)* and *val(\*,\*)* values are integer and not too large: *O(P(1)·…·P(d)·d·Q)* in the first case, and *O(P(1)·…·P(d-1)·d·Q)* in the second case.

# 4. ZERO- AND SINGLE-PLAYER GAME-THEORETIC MODELS

In this section we consider several zero- and single-player game-theoretic models, which are used for modeling the constraints of the resource management problems and in order to provide a context for the decisions of the players. Most games have one or more players and are studied from the perspective of the involved players (winning strategies, score maximization, and others). Zero-player games are somewhat unconventional and are sometimes not classified as games. The best known zero-player game is John Conway's Game of Life [9], which is described by a cellular automaton. Single-player games involve only one player, attempting to fulfill a game objective. This objective can be of two types: feasibility and optimization. In the first case, the player must reach a game state which belongs to a set of final states. In the second case, the player must reach a final state with the extra condition that the use of some resources is optimized (minimized or maximized).

## 4.1. A ZERO-PLAYER GAME BASED ON A LINEAR CELLULAR AUTOMATON

In this section we will consider a particular one-dimensional cellular automaton for which we will provide an algorithm which efficiently evaluates its state after any given number of time steps. The cellular automaton considered here consists of $n$ cells (numbered from $0$ to $n-1$) and, at any time moment, each cell can be in one of two states: $0$ or $1$. The automaton also has a transition function, which determines the state of each cell at the next time moment $t+1$, based on the states of the cell and those of its immediate neighbors at time moment $t$. If we denote by $q(i,t)$ the state of cell $i$ at time moment $t$, then $q(i,t+1)=f(q((i-1+n) \bmod n, t), q(i,t), q((i+1) \bmod n, t))$. Evaluating the state of each cell after one time step is easy, but evaluating it after a given number $m$ of time steps may require $O(n \cdot m)$ time. There are some classes of cellular automata, like linear additive automata, for which this evaluation can be performed in $O(n \cdot \log(m))$ time. The cells of a linear additive automata are positioned circularly (cell $n-1$ is the left neighbor of cell $0$) and their transition function is $q(i,t+1)=c_{-1} \cdot q((i-1+n) \bmod n, t)$ xor $c_0 \cdot q(i,t)$ xor $c_{+1} \cdot q((i+1) \bmod n, t)$, where $c_{-1}$, $c_0$ and $c_{+1}$ are constants from the set $\{0,1\}$. Because of the properties of the *xor* function, we have $q(i,t+2)=c_{-1} \cdot q((i-2+n) \bmod n, t)$ xor $c_0 \cdot q(i,t)$ xor $c_{+1} \cdot q((i+2) \bmod n, t)$ and, in general, $q(i,t+2^k)=c_{-1} \cdot q((((i-2^k) \bmod n) + n) \bmod n, t)$ xor $c_0 \cdot q(i,t)$ xor $c_{+1} \cdot q((i+2^k) \bmod n, t)$. With this, we can evaluate in $O(n)$ time the state of the automaton after $m=2^k$ steps. By writing $m=2^{p(1)}+2^{p(2)}+ \ldots +2^{p(r)}$, we can evaluate the state of the automaton in $O(n \cdot r)$ time, where $r=O(\log(m))$ (once we know the state of the automaton after $m'=2^{p(1)}+\ldots+2^{p(i)}$ steps, we can compute its state after $2^{p(i+1)}$ extra steps in $O(n)$ time, thus obtaining its state after $m''=m'+2^{p(i+1)}$ steps).

We will now consider a non-circular automaton, where, at each time step, every pair of adjacent cells $i \geq 0$ and $i+1<n$, such that $q(i,t)=1$ and $q(i+1,t)=0$ exchange their states (the $0$ and $1$ are swapped). The final state of such an automaton is reached after $T=O(n)$ steps, when all the $0$s are to the left of all the $1$s. A naive algorithm for computing the state of the automaton after every number $m \leq T$ of steps would take $O(n \cdot m)$ time. We will now provide an $O(n \cdot \log(n))$ algorithm for this problem. We will assign a number from $0$ to $nz-1$ to each zero cell of the automaton, in a left to right order ($nz$ is the total number of zero cells). The $i^{th}$ zero cell is initially located at the cell $c(i)$. It is obvious that all the zeroes *"move"* to the left and that, in the final (stable) state, the $i^{th}$ zero will be located at cell $i$. It is also obvious that the $i^{th}$ zero ($i \geq 1$) will not reach cell $i$ before the $(i-1)^{th}$ zero reaches cell $i-1$. During every time step, a zero cell performs an action: it either *"moves"* one cell to the left (if the state of the cell to the left is $1$) or *"waits"* (if the state of the cell to the left is $0$). For each zero cell $i$, we will determine the sequence of actions $a_{i,1}, a_{i,2}, \ldots, a_{i,na(i)}$ performed until it reaches its final cell (and the number of actions $na(i)$). The sequence will be maintained in reverse order, i.e. $a_{i,na(i)}$ is the action performed during the first time step and $a_{i,1}$ is the last action performed. Based on this sequence of actions, we will be able to determine in $O(\log(n))$ time the cell where each zero is located after $m$ time steps. Thus, in $O(n \cdot \log(n))$ time, we will determine the state of the automaton after any number $m$ of time steps. The number of time steps $T$ after which the final state is reached will be $T=na(nz-1)$. For the zero cell assigned number $0$, its sequence of actions consists of $na(0)=c(0)-0$ *"moves"*: $a_{0,j}=$*"move"* ($1 \leq j \leq na(0)$). We will determine the sequence of actions for each zero state, in increasing order of the assigned numbers. If $c(i)=c(i-1)+1$, then the sequence of actions for the $i^{th}$ zero cell is identical to the one for the $(i-1)^{th}$ zero cell, except that the first action performed is a *"wait"*. Thus, we have: $na(i)=na(i-1)+1$, $a_{i,j}=a_{i-1,j}$ ($1 \leq j \leq na(i)-1$) and $a_{i,na(i)}=$*"wait"*. If $c(i)>c(i-1)+1$, then the first $d=c(i)-c(i-1)-1$ actions of the $i^{th}$ zero cell will be *"moves"*. We need to find out if the $i^{th}$ zero cell *"catches up"* with the $(i-1)^{th}$ zero before the $(i-1)^{th}$ zero reaches its final cell and if it does, after how many time steps this situation occurs. If the $(i-1)^{th}$ zero cell performs less than $d$ *"waits"*, then the $i^{th}$ zero does not catch up with the $(i-1)^{th}$ zero and the actions performed by it will be: $a_{i,1}=a_{i,2}=\ldots=a_{i,c(i)-i}=$*"move"*. If the $i^{th}$ zero *"catches up"* with the $(i-1)^{th}$ zero after $t$ time steps (i.e. after $t$ time steps, it is located at the cell $x+1$, where $x$ is the cell where the $(i-1)^{th}$ zero is located), then we have $na(i)=na(i-1)+1$, $a_{i,na(i)}=$*"move"*, $\ldots$, $a_{i,1-t+na(i)}=$*"move"*, $a_{i,-t+na(i)}=$*"wait"* and $a_{i,-j+na(i)}=a_{i-1,-j+na(i)}$, for $t+1 \leq j \leq na(i)-1$.

In order to obtain the stated time complexity, we will maintain the actions in a compressed form: we will maintain an array *action*, where *action[j]="wait"* or *"move"*; we will also maintain an array *count*, where *count[j]* is *zero* if *action[j]="wait"*; *count[j]* denotes the number of consecutive *"move"* actions corresponding to *action[j]*. We will also maintain an array *totalCount*, where *totalCount[j]*=the sum of the values *count[1]+…+count[j]*. During the algorithm, we will also maintain two other arrays: *totalWaits[j]*=the number of *"wait"* actions in the (multi)set *{action[1], action[2], …, action[j]}* and *nextWait[j]*=the index (in the array *action*) of the next *"wait"* action performed after the action *action[j]* (if *action[j]="wait"* then *nextWait[j]=j*; else *nextWait[j]=nextWait[j-1]*). The algorithm below shows how to compute the sequence of actions and all the mentioned arrays efficiently. The array *action* is implemented as a stack. The sequence of actions corresponding to the $(i-1)^{th}$ zero cell is transformed into the sequence of actions of the $i^{th}$ zero cell. Similarly, all the other arrays are only transformed from the $(i-1)^{th}$ zero to the $i^{th}$ zero.

**LinearCellularAutomaton():**
*na=1; action[1]="move"; count[1]=$c_0$*
*totalWaits[0] = totalWaits[1] = nextWait[0] = nextWait[1] = 0*
*totalCount[0]=0; totalCount[1]=count[1]*
**for** *i=1* **to** *nz-1* **do** {
 **if** (*c(i)=i*) **then** *continue the for cycle*
 *t=d=c(i)-c(i-1)-1*
 **while** ((*na>0*) **and** (*d>0*)) **do** {
  **if** (*action[na]="wait"*) **then** {
   *d=d-1; t=t+1; na=na-1*
   **if** (*d=0*) **then** *break the while cycle* }
  **else** { // *action[na]="move"*
   *t=t+(totalCount[na]-totalCount[nextWait[na]])*
   *na=nextWait[na]* }}
 **if** (*na>0*) **then** {
  *na=na+1*
  *action[na]="wait"*
  *nextWait[na]=na; totalWaits[na]=totalWaits[na-1]+1*
  *count[na]=0; totalCount[na]=totalCount[na-1]* }
 *na=na+1*
 *action[na]="move"*
 *nextWait[na]=nextWait[na-1]*
 *totalWaits[na]=totalWaits[na-1]*
 *count[na]=t; totalCount[na]=totalCount[na-1]+t* }

At the end of each iteration of the outermost *for* cycle (*for i=1 to nz-1*), the arrays *action*, *nextWait*, *totalWaits*, *count* and *totalCount* contain the appropriate values for the $i^{th}$ zero cell. The values for the first zero cell (numbered with *0*) are available in the same arrays, before the first iteration of the outermost *for* cycle. Using these arrays, we can easily compute the cell that a zero state reaches after *m* time steps. If *m* is larger than the total number of actions performed *nact=totalCount[na]+totalWaits[na]*, then the zero state *i* is located at cell *i*. Otherwise, we need to compute the smallest index *j* (*1≤j≤na*), such that *nact-(totalCount[j-1]+totalWaits[j-1])≤m*. If *m=T*, we can find *j* in *O(1)* time (*j=1*); otherwise, we need to binary search the index *j*. Let *nw=totalWaits[na]-totalWaits[j-1]*, the number of *"waits"* performed during the first *m* time steps. Then, the $i^{th}$ zero is located at the cell *c(i)-m+nw*. This way, we can evaluate the state of the cellular automaton after *m* time steps in *O(n·log(n))* time (or *O(n)* time, if *m=T*). We can also compute *T*, as *max{totalCount[na]+totalWaits[na]}* after every iteration of the outermost *for* cycle (or before the first iteration).

**4.2. 1D PUSH-\***

*Push-\** is a simplified version of the well-known game *Sokoban*. A robot is placed in a 2D matrix consisting of unit squares which are either empty or contain a block. The robot can move in any of the four directions (if the corresponding square is free) and may also push blocks (any number of them) in a direction where an empty square exists. The purpose of the game is to bring the robot to a specified target square. In [8], 2D *Push-1* (pushing at most one block at a time) was proven to be NP-hard. In this section we consider the one-dimensional version of *Push-\**, with several additions. There are *N* squares on a linear board, numbered from *1* to *N* (from left to right). Some of the squares contain blocks, while others are empty. The robot starts in square *1* and must arrive at square *N*. In order to achieve this, the robot can make the following moves: *walk*, *jump* and *push*. A walk consists of moving from the current square *i* to the left (square *i-1*) or to the right (square *i+1*) if the destination square is empty (and without leaving the board). If the robot's square is *i* and square *i+1* contains a block, the robot may push that block one square to the right (together with all the blocks located between positions *i+2* and the first empty square to the right of *i+1*); obviously, at least one empty square must exist to the right of position *i+1* in order for the push to be valid. After the push, the robot's position becomes *i+1*. In a similar manner, the robot can push blocks to the left (if the square *i-1* contains a block, then all the blocks between position *i-1* and the first empty square to the left of square *i* are pushed one square to the left); after the push, the robot's position becomes *i-1*. The robot can also jump any number $Q$ ($1 \leq Q \leq K$) of squares to the right (left) if the previous $(K-1) \geq 1$ moves consisted of walking to the right (left). Each type of move consumes a certain amount of energy: *W* energy units for a walk, *P* units for a push and *J* units for a jump. In addition to reaching square *N*, the player should also do this by consuming the minimum total amount of energy. In the beginning, square *N* is empty and square *1* is occupied by the robot (thus, it contains no block).

We will find the minimum energy strategy with a dynamic programming approach. We will compute a table *E[i,j]*=the minimum energy consumed in order to have the robot located at square *i* and having *j* empty squares to the left (i.e., the squares *i-1*, *i-2*, ..., *i-j* are empty). Furthermore, the robot has not yet reached any square $k>i$ (thus, all the squares *i+1*, *i+2*, ..., *N* are in the same state as in the beginning). In order to justify the correctness of this approach, we will consider the squares grouped into intervals of consecutive empty squares. Let's number these intervals with consecutive numbers (starting from *1*), in the order in which they appear on the board (from left to right). If the robot reaches a square inside an interval *X*, then an optimal strategy will never contain moves which bring the robot to an interval $Y<X$. Thus, when the robot arrives in a square *i* inside an interval *X*, all the squares $k>i$ are in the initial state (have not been modified). This way, we can consider only sequences of moves which are local to the interval of consecutive empty squares into which the robot resides. The outcome of these moves should be that the player reaches another interval $Y>X$ (or another square $k>i$). We will show that for each state *(i,j)*, we need to consider only $O(N^2)$ sequences of moves, which will improve the value of some states *(i',j')*, with $i'>i$. Considering that there are $O(N^2)$ possible states, the time complexity of the algorithm will be $O(N^4)$. We will first compute an array *dmin*, where *dmin[d]*=the minimum energy needed to travel d squares. We have that:

$$\text{dmin}[1] = W$$

$$\operatorname*{dmin}_{2 \leq d \leq N}[d] = \min \begin{Bmatrix} \text{dmin}[k] + \text{dmin}[d-k], 1 \leq k < d \\ W \cdot \lfloor d/2 \rfloor + J \end{Bmatrix} \quad (1)$$

We will also compute the following arrays: *next*, where *next[i]*=the next square to the right of square *i*, which contains a block (if no such square exists, then *next[i]=N+1*; *next[N]=N+1* and *next[1≤i≤N-1]=if square i+1 contains a block then i+1 else next[i+1]*), *nbleft*, where *nbleft[i]*=the number of blocks on the squares *1,2,...,i*, and *nbright*, where *nbright[i]*=the number of blocks on the squares *i, i+1, ..., N*. These arrays can be computed in *O(N)* time each. Initially, we have *E[1,0]=0* and *E[i,j]=+∞* (for *i>1* or *j>0*). From each state *(i,j)*, such that *E[i,j]<∞*, we will consider several types of moves, depending on the value of *next[i]*. If *next[i]=N+1*, then we will consider the following moves:
 1. move *x* squares to the left and then travel *y* squares to the right.
 2. move *j* squares to the left, push *x* squares to the left and then travel *y* squares to the right.

If *next[i]≤N*, then we consider the following moves:
 1. travel *x* squares to the right.
 2. walk *x* squares to the right and then jump *y* squares to the right.
 3. travel *x* squares to the left, then walk *next[i]-i+x-1* squares to the right and jump *y* squares to the right
 4. travel *j* squares to the left, push *x* squares to the left, walk *next[i]-i+j-1* squares to the right and then jump *y* squares to the right.
 5. travel *next[i]-i-1* squares to the right, push *x* squares to the right, travel *y* squares to the left, walk *y* squares to the right and jump *y+1* squares to the right.
 6. travel *next[i]-i-1* squares to the right, push *x* squares to the right, travel *next[i]+x-i+j* squares to the left, push *y* squares to the left, walk *next[i]+x-i+j+y* squares to the right and then jump *next[i]+x-i+j+ y+1* to the right.

When pushing *x* squares to the left, we need to make sure that there are at least *x* empty squares available to the left, i.e. *i-j-nbleft[i]≥x*. When pushing *x* squares to the right, we need to have *N-next[i]+1-nbright[i]≥x*. When jumping *x* squares to the right, the landing square *i'* must be empty (both in the initial state and after performing the sequence of moves) and the value of *x* must be at most *K*, where $K-1 \geq 1$ is the number of consecutive walks to the right performed right before the jump. Every sequence of moves ends with a jump to the right (for the case *next[i]≤N*). The jump makes sure that the robot moves to a different interval of consecutive empty squares. We need to determine the state *(i', j')* reached by the player after the sequence of moves. It is easy to determine the landing square *i'*. We also have to find out the number *j'* of consecutive empty squares directly to the left of *i'*. This number might be the same as in the initial state, or smaller, because of the possible right pushes performed during the sequence of moves. For each square *i*, we will compute *neleft[i]*=the number of consecutive empty squares (in the initial state) immediately to the left of square *i* (i.e.

squares *i-1, i-2, …, i-neleft[i]* are empty and square *i-neleft[i]+1* contains a block or is outside of the board) and *nteleft[i]*=the total number of empty squares in the set *{1,2,…,i}*. Before performing a sequence of moves, there will be *ne=nteleft[i']-nteleft[next[i]]* empty squares between *next[i]* and *i'*. If the sequence of moves contained $x \geq 0$ pushes to the right, then we distinguish the following cases:

- *x>ne*: square *i'* is not empty and, thus, the robot cannot land there after the jump
- *x≤ne-neleft[i']*: square *i'* has *j'=neleft[i']* consecutive empty squares immediately to its left
- *ne-neleft[i']<x≤ne*: square *i'* has *j'=ne-x* consecutive empty squares immediately to its left

If the energy consumed by performing the sequence of moves is *ES*, then we need to set *E[i',j']* to *min{E[i',j'], E[i,j]+ES}*. We notice that each sequence of moves contains at most two variable parameters (*x* and *y*). There are $O(N)$ possible values for *x* and *y* (starting from *0*) and, thus, there are $O(N^2)$ possible sequences of moves for each type of move. The algorithm's time complexity is, thus, $O(N^4)$.

**4.3. RESOURCE COLLECTOR 1**

Let's consider a complete directed graph with *N* vertices, numbered from *1* to *N*. The player is initially located in vertex *1*. For each ordered pair of vertices *(i,j)*, the time required to travel from *i* to *j* on the shortest path, $tr_{i,j}$, is known ($tr_{i,i}=0$). At certain time moments, recipients with resources may appear in the vertices of the graph. Considering that there are *M* recipients overall, for each recipient *k*, the time moment when it appears, $ta_k$, the vertex where it appears, $v_k$, and the quantity of resources in the recipient, $c_k \geq 0$, are known (if multiple recipients appear at the same time and at the same vertex, we will replace them by a single recipient whose quantity of resources is equal to the sum of the resources in the initial recipients). All the time moments are considered to be integers. At each moment *t*, the player may either stay in its current position (vertex *i*) or may start traveling towards another vertex *j* (which he/she reaches at time moment $t+tr_{i,j}$). The resources inside a recipient *k* can be collected by the player only if the player is located at vertex $v_k$ at the moment the recipient appears ($ta_k$) or if the player just arrives at the vertex at that moment. The purpose of the game is to collect the largest possible quantity of resources (knowing in advance all the parameters).

An optimal strategy can be found by using dynamic programming. We sort the recipients in increasing order of their moment of appearance (breaking ties arbitrarily). Thus, recipient *k* appears after (or at exactly the same time as) every recipient *p<k*. For each recipient *k*, we will compute $C_{max}[k]$=the maximum quantity of resources which the player can collect if at time $ta_k$ it arrives (or is already located) at vertex $v_k$ (and, thus, collects the resources in recipient *k*). We will also consider a virtual recipient with *0* resources, appearing at vertex *1* at time moment *0* (this recipient is assigned number *0*). We have $C_{max}[0]=0$ and for $k \geq 1$:

$$C_{max}[k] = c_k + \max \begin{cases} C_{max}[p], \text{if } 0 \leq p < k \text{ and } tr_{v_p, v_k} \leq ta_k - ta_p \\ -\infty \end{cases} \quad (2)$$

The maximum quantity of resources which can be collected is the maximum value in the array $C_{max}$. The time complexity of this algorithm is $O(M^2)$ and it is efficient only when the number of recipients is not too large. We will now present some efficient algorithms for the case when *M* is very large: for instance, *M>N* and/or $M>T_{max}$, where $T_{max}$ is an upper limit for the maximum travel time between any two vertices.

We will compute the same values as above, but we will make the following observation: if $v_p=v_k$ (*p<k*), then $C_{max}[p] \leq C_{max}[k]$. Indeed, $C_{max}[k]$ could be obtained, for instance, by collecting the resources in the recipient *p* and then waiting at vertex $v_p$ until the time moment $ta_k$ (if no better strategy exists). For each vertex *i*, we will maintain a list with all the recipient numbers which appeared at vertex *i*, sorted in chronological order. Let this list be *cb(i,1), cb(i,2), …, cb(i,ncb_i)*, where $ncb_i$ is the number of recipients which appeared at vertex *i* (so far). When computing $C_{max}[k]$ for a recipient *k*, we will iterate over all the vertices of the graph. For each vertex *i*, we will find the last recipient *cb(i,j)*, such that $tr_{i,v_k} \leq ta_k - ta_{cb(i,j)}$ and set $C_{max}[k]=max\{C_{max}[k], c_k+C_{max}[cb(i,j)]\}$. Since the recipients *cb(i,1), …, cb(i,ncb_i)* are sorted such that $ta_{cb(i,1)} < … < ta_{cb(i,ncb_i)}$, we can perform a binary search in order to find the recipient *cb(i,j)*. Thus, the time complexity becomes $O(M \cdot N \cdot log(M))$. After computing $C_{max}[k]$, we add *k* at the end of the list of recipients of the vertex $v_k$.

When the maximum travel time between any two vertices *i* and *j* ($tr_{i,j}$) is less than (or equal to) a small value $T_{max}$, we can improve the algorithm further. For each vertex *i*, we will maintain a value $T_{last}[i]$=the most recent time moment when a recipient appeared at vertex *i*. We will also maintain a table *MaxC[i,t]*, with $0 \leq t \leq T_{max}$, representing the maximum quantity of resources the player can gather if at time $T_{last}[i]-t$ he/she is located at vertex *i*. Initially, $T_{last}[i]=0$, for all the vertices *i*, and $MaxC[i,t]=-\infty$, except for *MaxC[1,0]*, which is *0*. With these values, we will compute $C_{max}[k]$ using the algorithm presented next, whose time complexity is $O(M \cdot (N+T_{max}))$. Afterwards, we will consider a situation in which the graph corresponds to a geometric arrangement of the vertices.

**ResourceCollector1-SmallValueOfTmax:**

```
for k=1 to M do {
 C_max[k]=-∞
 for i=1 to N do {
  t_latest=ta_k-tr_{i,vk}
  if (T_last[i]<t_latest) then nc=MaxC[i,0]
  else nc=MaxC[i,T_last[i]-t_latest]
  C_max[k]=max{C_max[k], c_k+nc} }
 // update the values MaxC[v_k,t] and T_last[v_k]
 if (ta_k-T_last[v_k]>T_max) then {
  t_offset=T_max+1 }
 else { t_offset=ta_k-T_last[v_k] }
 nc=MaxC[v_k,0]
 for t=T_max down to t_offset do MaxC[v_k,t]=MaxC[v_k,t-t_offset]
 for t=0 to t_offset-1 do MaxC[v_k,t]=nc
 T_last[v_k]=ta_k; MaxC[v_k,0]=C_max[k] }
```

If the graph's vertices are points on the *OX* axis (each point *i* having a coordinate $x_i$) and the travel times between two vertices *i* and *j* is the difference between their coordinates ($tr_{i,j}=|x_i-x_j|$), we can improve the time complexity of the solution. We consider a two-dimensional plane, in which the *OX* axis corresponds to the coordinates of the vertices and the *OY* axis corresponds to time moments. With this representation, each recipient *k* is a point with coordinates $(x_{vk}, ta_k)$. When computing the value $C_{max}[k]$ of the recipient *k*, we are interested in the $C_{max}$ values of the recipients *p* ($0\leq p<k$) whose coordinates have the following property: $|x_{vp}-x_{vk}|\leq ta_k-ta_p$. This condition defines a rectangular quarter-plane, with the origin in $(x_{vk}, ta_k)$. The quarter-plane is rotated *45* degrees from the orientation of the *OX* and *OY* axes. By rotating all the points associated to the recipients by *-45* degrees around the origin, each recipient is assigned some new coordinates $(x_k', y_k')$. With the new coordinates, the condition for a recipient *p<k* to be considered when computing $C_{max}[k]$ is: $x_p'\leq x_k'$ and $y_p'\leq y_k'$. The quarter-plane is now aligned with the *OX'* and *OY'* axes. If we consider the value $C_{max}[k]$ of a recipient *k* to be the weight of the point $(x_k', y_k')$, we are interested in finding the maximum weight of a point located inside a rectangle for which two sides are unbounded (quarter-plane with a corner at a given point). We can use orthogonal range search results for solving this problem. We need to consider the dynamic version of the orthogonal range maximum query problem, however, because the weights of the points can change (initially, the weights are -∞ and they change at the moments when the $C_{max}$ values are computed). An orthogonal range query and update can be performed in $O(log^2(M))$ time, using a 2D range tree. Each node of the range tree is assigned an interval of x-coordinates and stores all the points with the x-coordinates inside the assigned interval. The space requirement is $O(M \cdot log(M))$. The points stored at each node are inserted into a balanced binary tree, whose search key is given by the y-coordinates. Each node of the balanced tree maintains the maximum weight of a point inside its subtree. A query partitions the x-interval into $O(log(M))$ sub-intervals corresponding to $O(log(M))$ range tree nodes. For each tree node, its corresponding balanced tree is searched and the maximum weight of a point belonging to the query y-interval is found. An update removes a point from the balanced tree of each range tree node to which it belongs and reinserts it with the new weight. Thus, the algorithm has $O(N+M \cdot log^2(M))$ time complexity.

### 4.4. RESOURCE COLLECTOR 2

We have $N\geq 4$ recipients (numbered from *1* to *N*); recipient *i* ($1\leq i\leq N$) contains $r(i)\geq 0$ resources units. We want to develop a strategy which collects all the resources into one single recipient, using the following type of actions: *Move(u,v,w)*=choose three distinct recipients *u*, *v* and *w*, such that $r(u)>0$ and $r(v)>0$, and then decrease $r(u)$ and $r(v)$ by *1* each, and increase $r(w)$ by *2* (we effectively move *1* unit of resource from both recipients *u* and *v* to recipient *w*). At first, we check if all the resources are already gathered in only one recipient (i.e. if there are at least *N-1* distinct recipients *i* with $r(i)=0$). Afterwards, we also check for the only case which cannot be solved: when we have only two recipients *u* and *v* with non-zero resources, and $r(u)=2$ and $r(v)=1$. If we are not in any of these two cases, we will use a two-stage algorithm with $O(N+M)$ time complexity, which we present next. *M* is the total number of moves and will be proportional to the sum $r(1)+r(2)+...+r(N)$.

During the first stage, our goal will be to move all the resources to recipient *N*. We initialize *i=j=1*. Then, while (*i<N*) and (*j<N*), we perform the following steps: *(1)* while ($r(i)=0$) and (*i<N*), we increase *i* by *1*; *(2)* while (($j\leq i$) or ($r(j)=0$)) and (*j<N*), we increase *j* by *1*; *(3)* if (*i<j*) and (*j<N*) then we perform a move *Move(i,j,N)* (i.e. we decrease $r(i)$ and $r(j)$ by *1* each, and then we increase $r(N)$ by *2*). After this initial stage, we have two possible outcomes: *(1)* all the resources were gathered in recipient *N*; *(2)* there is exactly one more recipient *k* ($1\leq k\leq N-1$) with $r(k)>0$. We search for *k* in $O(N)$ time. If we find a recipient *k<N* with $r(k)>0$, then we are in case *2*. We will now choose two recipients *a* and *b*, distinct from *k* and *N* (e.g. if *k=1*, then *a=2* and *b=3*; if *k=N-1*, then *a=N-2* and *b=N-3*; if $2\leq k\leq N-2$, then *a=k-1* and

$b=k+1$). Then, if $r(k)>r(N)$, we set *dest=k* and then *k=N*; otherwise, *dest=N*. *dest* is the index of the recipient where we intend to eventually gather all the resources and *k* is the index of the other recipient which still contains non-zero resources. While $(r(k)>0)$ and $(r(dest)>0)$ we perform the following actions: *(1)* if $r(k) \geq 2$ then we perform the moves *Move(k, dest, a)*, *Move(k, dest, b)* and then twice the move *Move(a, b, dest)*, in this order (after this sequence of moves, *r(k)* is effectively decreased by *2* and *r(dest)* is increased by *2*, while *r(a)* and *r(b)* remain zero); *(2)* otherwise (if $r(k)=1$), we perform the move *Move(k, dest, a)*, and then we swap the values of *k* and *a* (i.e. *vaux=k*, *k=a*, and then *a=vaux*). In the end, all the resources will be gathered either in the recipient *dest* or, if we reached the case $r(k)=r(dest)=1$, then they will be gathered in the recipient *k*.

## 4.5. ORDERING TOKENS BY MOVING PAIRS OF ADJACENT TOKENS

We have a board consisting of $L=2 \cdot N+2$ positions (numbered from *1* to *L*), with $N \geq 3$. Each position *i* ($1 \leq i \leq L$) is occupied either by a token of color *B*, a token of color *R*, or is empty. There are *N* tokens of each color placed on the board and, thus, two positions are empty. The two empty positions are adjacent. We can perform the following type of moves: *Move(i)*=we move the tokens on the positions *i* and *i+1* to the positions *p* and *p+1*, where *p* and *p+1* are the two empty positions (the token on position *i* is moved to position *p* and the token on position *i+1* is moved to position *p+1*); positions *i* and *i+1* must necessarily contain a token each (of any and possibly different colors); as a result of this move, the two new empty positions will be *i* and *i+1*. We want to perform a sequence of moves such that, at the end, the positions *1, ..., N* are occupied by the tokens of color *R*, the positions *N+1* and *N+2* are empty, and the positions *N+3, ..., 2·N+2* are occupied by the tokens of color *B*.

A first solution would be to encode every possible state of the board as a base *3* number (with a value *0* for a position containing a token with color *R*, a value *1* for a position containing a token with color *B*, and a value *2* for an empty position). Since every such encoding contains exactly two empty positions, not all the numbers with *L* digits in base *3* are valid; however, the number of valid base *3* encodings is exponential in the number of positions of the board. We can then construct a graph of the encodings (where the encodings are vertices). We add an edge between an encoding *A* and an encoding *B* if there is a move such that the board changes from the configuration corresponding to the encoding *A* to the configuration corresponding to the encoding *B*. Thus, if *S* is the encoding of the initial state of the board and *D* is the encoding of the final state of the board, we just need to find a path from *S* to *D* in this graph (we can do this with a simple breadth-first search). Then, by following the path from *S* to *D*, we know exactly which moves need to be made (and in which order). The problem with this approach is that the time complexity is exponential in the parameter *L*.

We will now present an algorithm whose time complexity is $O(N)$. At first, we search the board and find the leftmost empty position *p* (i.e. positions *p* and *p+1* are empty). During the algorithm, after performing every move *Move(z)*, we will set *p=z* (although we will not explicitly mention this in the description of the algorithm). We will initialize a counter *i=1*. This counter will have the property that on the positions *1, ..., i-1*, we will only have tokens colored with *R*. We will also maintain a counter *nextR* (initially *0*), representing the position of the next token with color *R*, located after the position *i*. Then, while ($i \leq N$), we perform the following actions: *(1)* if the position *i* contains a token with color *R*, then we just set *i=i+1* and then we continue; *(2)* otherwise, if the position *i* contains a token with color *B*, we have two sub-cases: *(2.1)* if the position *i+1* is not empty then we perform the move *Move(i)* (and we do not change the counter *i*; however, at the next iteration, the position *i* will be empty); *(2.2)* otherwise, if position *i+1* is empty (i.e. *p=i+1*), we perform the move *Move(i+3)*, i.e. we move the tokens from the positions *i+3* and *i+4* to the empty positions *i+1* and *i+2* (again, we do not change the value of the counter *i*, but at the next iteration, the position *i+1* will not be empty anymore); *(3)* if, at the beginning of the iteration, the position *i* is empty (i.e. *i=p*), then: *(3.1)* if ($nextR \leq i$) then we set *nextR=i+1*; *(3.2)* while the position *nextR* does not contain a token with color *R*, we increment *nextR* by *1*; *(3.3.1)* if ($nextR<L$) then we perform the move *Move(nextR)* (without changing the value of the counter *i*; however, at the next iteration, position *i* will contain a token colored with *R*, and, thus, the counter *i* will be subsequently incremented by *1*); *(3.3.2)* if ($nextR=L$) then we reached the last token colored with *R*: this means that *i=N* and that all the other *N-1* tokens colored with *R* are on the positions *1, 2, ..., N-1*; in this case, we will perform the following sequence of moves: *Move(L-1)* (after this move, position *N* contains a token with color *B* and position *N+1* contains a token with color *R*; the empty positions are *L-1* and *L*), *Move(N+2)* (after this move, the positions *L-1* and *L* each contain one token with color *B* and the positions *N+2* and *N+3* are empty), *Move(N-1)* (after this move, the positions *N+1* and *N+2* each contain one token with color *R* and the positions *N-1* and *N* are empty), *Move(N+1)* (after this final move, all the positions *1, 2, ..., N* contain tokens colored with *R*, the positions *N+1* and *N+2* are empty, and the positions *N+3, ..., 2·N+2* each contain a token colored with *B*). At the end of the algorithm, the positions *1, ..., N* contain tokens colored with *R*. If the empty positions *p* and *p+1* are not *N+1* and *N+2*, then: *(1)* if $p>N+2$ then we call *Move(N+1)*; *(2)* if ($p=N+2$) then we first call *Move(N+4)* and, then, *Move(N+1)*.

# 5. GEOMETRIC OPTIMIZATION PROBLEMS

Geometric optimization problems arise in the resource management field, because resources are commonly modeled as points in a multidimensional feature space, in which every dimension corresponds to an attribute of the resource. In this section we consider two such problems: a geometric aggregate coverage problem and the hyper-rectangle k-center problem using the $L_\infty$ metric.

## 5.1. GEOMETRIC MINIMUM AGGREGATE COVERAGE

We are given *n* rectangles in the plane ($[xa(i,1),xb(i,1)]$ x $[xa(i,2),xb(i,2)]$, $xa(i,j) \leq xb(i,j)$, $1 \leq j \leq 2$, $1 \leq i \leq n$) located inside a larger rectangle *R* ($[xaR(1),xbR(1)]$ x $[xaR(2), xbR(2)]$, $xaR(j) \leq xbR(j)$, $1 \leq j \leq 2$). Each rectangle *i* ($1 \leq i \leq n$) has a weight $w(i)>0$. We want to place inside *R* a rectangle *R'* of side lengths *L(1)*, *L(2)* ($L(j)>0$ is the side length in dimension *j*, $1 \leq j \leq 2$). The cost of placing *R'* at a given position is given by an aggregate function over the weights of the rectangles intersected by *R'*. The aggregate function can be +, ∗ or *max*. We want to find a placement of *R'* inside *R* whose cost is minimum. We will first handle the case where the aggregate function is +. The location of *R'* is completely specified by the coordinates of its upper-right corner. We will inflate each of the n rectangles, by extending the length in dimension *j* by *L(j)*. To be more precise, we set $xb(i,j)=xb(i,j)+L(j)$, $1 \leq j \leq 2$, $1 \leq i \leq n$. Now we can redefine the problem as follows. We want to find a point inside the rectangle *R* with minimum placement cost, where the placement cost is equal to the aggregate of the weights of the rectangles containing the point. We will solve this problem as follows. At first, we « clip » every coordinate $xb(i,j)$ such that it doesn't exceed $xbR(j)$, i.e. we set $xb(i,j)=min\{xb(i,j), xbR(j)\}$. Next, we will sort all the *2·n+2* coordinates of the *n* rectangles, plus the large rectangle *R*, in each dimension. Let $xso(j,1) \leq xso(j,2) \leq ... \leq xso(j,2 \cdot n+2)$ be the sorted coordinates in dimension *j* (j=1,2). We will remove all the duplicates, obtaining *m(j)* distinct coordinates in each dimension *j*, $xs(j,1)<xs(j,2)<...<xs(j,m(j))$. For each rectangle *i* ($1 \leq i \leq n$) we compute *idxa(i,j)*=the index of the coordinate *xa(i,j)* in the sorted list *xs(j)*, i.e. $xs(j,idxa(i,j))=xa(i,j)$. We can do this by using binary search. Similarly, we compute *idxb(i,j)*, such that $xs(j,idxb(i,j))=xb(i,j)$. We will sweep the rectangles from left to right (in the first dimension) and for each entry *p* in the 2$^{nd}$ dimension ($1 \leq p \leq m(2)$) we will maintain a cost *cost(p)*, representing the cost of placing a rectangle with the upper side at coordinate *xs(2,p)*. In order to efficiently maintain the values *cost(p)*, we will construct a segment tree over the *m(2)* (distinct) coordinates in the 2$^{nd}$ dimension. The coordinate values are not important, only their index in the list of sorted coordinates. Each node of the segment tree will maintain two values : *qagg*, the query aggregate over the indices in its range and *uagg*, the update aggregate of all the updates which « stopped » at that node. For more explanations and a comprehensive algorithmic framework using segment trees, see [3]. During the sweep, we will have two types of events : we encounter the left side of a rectangle or we encounter the right side of a rectangle. The events will be sorted according to their coordinates in the first dimension. In case of multiple events at the same coordinate, we will consider first the *right side* events, followed by the *left side* events for that coordinate. When a left side for a rectangle *i* occurs, we update the interval *[idxa(i,2), idxb(i,2)-1]*, by adding the value *w(i)* to all the values *cost(p)*, with *p* inside the interval. An update will be performed in *O(log(n))* time, by computing a canonical decomposition of the interval, consisting of *O(log(n))* segment tree nodes. We increase the *uagg* values of these tree nodes by *w(i)*. Then, if the node is a leaf, we also increase *qagg* by *w(i)* ; otherwise, we set *qagg* to *uagg+min{qagg(leftson(node)), qagg(rightson(node))}*. Then, we recompute the *qagg* values of all the *O(log(n))* ancestors of the tree nodes which are part of the canonical decomposition. For each ancestor node *a*, we set *qagg(a)* to *uagg(a)+min{qagg(leftson(a)), qagg(rightson(a))}*. This way, the *qagg* value of the tree root will always be equal to the minimum value of *cost(p)* ($1 \leq p \leq m(2)$) at the current position. If we encounter a right side event, we perform the same set of actions as before, except that we add -*w(i)* instead of *w(i)* at the *uagg* values of the tree nodes of the canonical decomposition of the (same) interval. After every update, we compare the *qagg* value of the tree root against the minimum cost found so far and update this cost, if the *qagg* value is smaller. Thus, in *O(n·log(n))* time, we can compute the minimum cost. If we also maintain the coordinate of the current event as well as the index *p* for which the minimum value is attained, we can also find where to place the rectangle.

The case where the aggregation function is ∗ is identical to the + case. We replace every weight *w(i)* by *log(w(i))*. The multiplication is known to be equivalent to the addition of the logarithms. The position achieving the minimum value for the logarithms case with + as the aggregation operator also achieves the minimum value in the normal weights case, with ∗ as the aggregation operator.

In order to support the *max* aggregation operator, each *uagg* value will be replaced by a balanced tree. For each left side event, we insert the value *w(i)* in the balanced trees of all the nodes of the canonical decomposition. For each leaf node in the decomposition we set *qagg* to *uagg.getMax()*. For the other nodes of the canonical decomposition and for the ancestors of the nodes in the decomposition (in this order) we set *qagg* to *max{uagg.getMax(), min{qagg(leftson(node)), qagg(rightson(node))}}*. For a right side event, we remove *w(i)* from the balanced trees *uagg* of the nodes in the decomposition and recompute the *qagg* values as before. If *uagg* contains no values, then *uagg.getMax()* returns *0* ; otherwise, it returns the largest value in *uagg*. The time complexity in this case is $O(n \cdot log^2(n))$, but we use $O(n \cdot log(n))$ memory storage, because every weight of a rectangle is stored in $O(log(n))$ tree nodes. We can also solve the problem with $O(n)$ storage. We sort the weights of all the rectangles and then we binary search the minimum weight $W_{min}$ of a rectangle that cannot be avoided when placing the rectangle *R'*. The feasibility test for a candidate value $W_{cand}$ starts by ignoring all the rectangles *i* with $w(i)<W_{cand}$. For the remaining rectangles we apply the same transformations (we inflate them) and arrive at the problem of finding the location of a point which is not contained in any of the inflated rectangles. This problem was considered in [2] and solved in $O(n \cdot log(n))$ time with $O(n)$ storage. The overall complexity is $O(n \cdot log^2(n))$.

If we want to find the largest rectangle *R'* with a fixed aspect ratio, i.e. *L(2)=f·L(1)* (where *f* is a constant), such that the placement cost is at most *B*, we can binary search the length *L(1)* and compute the minimum placement cost for every candidate value. If the cost is at most *B*, we can test a larger value ; otherwise, we will test a smaller one. This approach adds an *O(log(LMAX))* factor to the time complexity (*LMAX* is the length of the search interval for *L(1)*).

## 5.2. D-DIMENSIONAL HYPER-RECTANGLE K-CENTERS

We are given *n* (unweighted) points in *d*≥*2* dimensions; point *i* is located at coordinates *(x(i,1),…,x(i,d))*. We want to place *K* hyper-rectangles with side lengths *L(i,j), 1≤i≤K, 1≤j≤d* (*L(i,j)* is the side length in dimension *j* of the hyper-rectangle *i*) such that the maximum ($L_\infty$) distance from a point to the closest hyper-rectangle is minimized (the distance is *0* if the point is contained inside a hyper-rectangle). We binary search the maximum distance *D* and *inflate* each hyper-rectangle *i* to *L'(i,j)=L(i,j)+2·D (1≤j≤d)*. The feasibility test consists of verifying if all the points can be covered by the *K* hyper-rectangles of sizes *L'(i,j), 1≤i≤K, 1≤j≤d*. We consider the following recursive function, given in pseudocode, which verifies if a set of points *S* can be covered by the first *K* identical hyper-rectangles with side lengths *L(i,j)* (*1≤i≤K*) in dimension *j* (*1≤j≤d*).

**IsFeasibleCover(d, S, K):**
**if** *(|S|=0)* **then return** *"feasible"*
**if** *(K=1)* **then {**
  **for** *j=1* **to** *d* **do {**
    *xmin(j)=min{x(i,j)|i is a point in S}*
    *xmax(j)=max{x(i,j)|i is a point in S}* **}**
  **if** *(xmax(j)-xmin(j)≤L(1,j)* **for each** *1≤j≤d)* **then return** *"feasible"* **}**
**else {** // K≥2
  **for** *j=1* **to** *d* **do {**
    *dc(j)=the number of distinct coordinates x(i,j), i∈ S (sort the coordinates in dimension j of all the points in S and remove the duplicates)*
    **let** *co(j, 1) < … < co(j, dc(j)) =* the distinct coordinates in dimension *j*
    **compute** *comin(j)=min{i |1≤i≤dc(j), co(j,i)+L(K,j)≥co(j,dc(j))}* **}**
  **if** *(co(j,dc(j))-co(j,1)≤L(K,j)* **for each** *1≤j≤d)* **then return** *"feasible"*
  $q = \lfloor 2 \cdot d / K \rfloor$
  **if** *(K=2)* **then {**
    **for** *j=1* **to** *d* **do {**
      **let** *RT(j)=a d-dimensional range tree*
      **insert** *into RT(j) every point i∈ S, with weight w(i)=x(i,j)* **}}**
  **for each** *subset SD of {-1,-2, …, -d, 1,2,…, d} such that |SD|=q* **do {**
    **for each** *tuple (c(1), …, c(d)), such that (1≤c(j)≤dc(j), 1≤j≤d) and (consistent(c(1), …, c(d), SD)=true)* **do {**
      *consider a hyper-rectangle R with 2 opposite corners at (co(1,c(1)), …, co(d,c(d))) and (co(1,c(1))+L(K,1), …, co(d,c(d))+L(K,d))*
      **if** *(K=2)* **then {**
        **for** *j=1* **to** *d* **do {**
          *xmin'(j)=+∞; xmax'(j)=-∞*
          **for** *j'=1* **to** *d* **do {**
            *xmin'(j)=min{xmin'(j), RT(j).getMinWeight(x(\*,j') < co(j',c(j'))), RT(j).getMinWeight(x(\*,j')>co(j',c(j'))+L(K,j'))}*
            *xmax'(j)=max{xmax'(j), RT(j).getMaxWeight(x(\*,j') < co(j',c(j'))), RT(j).getMaxWeight(x(\*,j')>co(j',c(j'))+L(K,j'))}* **}}**
        **if** *(xmax'(j)-xmin'(j)≤L(K-1,j)* **for each** *1≤j≤d)* **then return** *"feasible"* **}**
      **else {** // K≥3
        *S'={i| i∈ S, i is contained in R}*
        **if** *(***IsFeasibleCover***(d, S\S', K-1)) then return "feasible"* **}}}}**
**return** *"not feasible"*
**consistent(c(1), …, c(d), SD):**
**for** *j=1* **to** *d* **do {**

  **if** *((-j)∈ SD)* **and** *(c(j)>1))* **then return false**
  **if** *((j∈ SD)* **and** *(c(j)<comin(j)))* **then return false** }
**return true**

    For $K=1$, the problem can be easily solved in $O(n)$ time, by verifying if the minimum bounding box (MBR) of the points can be included in the hyper-rectangle. For $K \geq 2$, we compute the value of $q$, having the following meaning: hyper-rectangle $K$ must have at least $q$ of its $2 \cdot d$ sides along $q$ sides of the MBR of the points in $S$. Thus, we consider all the $C(2 \cdot d, q)$ possibilities of choosing these sides ($C(a,b)$=combinations of $a$ elements taken $b$ at a time). For each possibility, we consider all the coordinates for the leftmost corner (in each dimension) of the $K^{th}$ hyper-rectangle which are consistent with the chosen sides. For each such possibility, we verify if we can cover the remaining points with $K-1$ hyper-rectangles. When $K \geq 3$, we perform this test by calling the recursive function with the set of yet uncovered points and $K-1$ as parameters. When $K=2$, we could do the same (calling the function for $K=2-1=1$). However, by maintaining a range tree $RT(j)$ for every coordinate $j$, in which we insert all the points in the set $S$ with weights equal to their $x(*,j)$ coordinate, we can compute the coordinates of the MBR of the remaining points in $O(d^2 \cdot log^d(n))$. The range tree provides two functions: *getMinWeight(r)* and *getMaxWeight(r)*, where $r$ is a d-dimensional orthogonal range, which return the minimum (maximum) weight of a point in the tree in the given orthogonal range; if no point exists in that range, then they return $+\infty$ ($-\infty$). The ranges we consider are of the form $(x(*,j)>a)$ and $(x(*,j)<a)$, i.e. $(a,+\infty)$ $((-\infty,a))$ in dimension $j$ and $(-\infty,+\infty)$ in the other dimensions. For $K=2$ (and $d$ constant), the time complexity is $T(n,m,K=2)=O(n^{d/2} \cdot log^d(n))$, where $m$ is the maximum number of distinct coordinates in any dimension. For $K \geq 3$, we have $T(n,m,K)=n+O(C(2 \cdot d, 2 \cdot d/K)+m^d) \cdot T(n,m,K-1)=O(n^{(K-2) \cdot d+d/2} \cdot log^d(n))$.

    The $(K+P)$-center problem is an extension of the K-center problem and considers that $P$ fixed hyper-rectangles (not necessarily identical with each other or with some of the other $K$ which we need to place) are given at no cost. The corresponding $(K+P)$-center problem is solved by inflating accordingly the $P$ fixed hyper-rectangles and ignoring in the decision problem all the points which are contained in at least one of the $P$ inflated hyper-rectangles. We can easily find all the points contained in at least one of the $P$ inflated hyper-rectangles in $O(n \cdot P)$ time, but, if $P$ is large, we can do this in $O(P \cdot log(n)+n \cdot log(n))$. We construct a (dynamic) d-dimensional range tree with all the $n$ points. Then, for each of the $P$ hyper-rectangles, we report in $O(log(n)+qp)$ time, all the $qp$ points contained in it; afterwards, we remove the $qp$ points from the range tree (in order to avoid reporting the same point multiple times, which would again take $O(n \cdot P)$ time).

    When the points may have different weights and all the $K$ hyper-rectangles are identical (their side lengths are $L(1)$, …, $L(d)$), the hyper-rectangle $K$-center problem is equivalent to deciding if a set of $n$ hyper-rectangles is $K$-pierceable. We binary search the minimum weighted distance $D$. For a candidate value $D_{cand}$, we assign a d-dimensional hyper-rectangle to every point $i$: $HR_i=[x(i,j)-L(j)-D_{cand}/w(i), x(i,j)+D_{cand}/w(i)]$ $(1 \leq j \leq n)$. We must have the lower corner of one of the hyper-rectangle-centers within every hyper-rectangle $HR_i$. Thus, $D_{cand}$ is a feasible weighted distance if the hyper-rectangles $HR_i$ are K-pierceable. The case $K=1$ is easy, as it is equivalent to deciding if $max\{x(i,j)-L(j)-D_{cand}/w(i)|1 \leq i \leq n\}$ $\leq min\{x(i,j)+D_{cand}/w(i)|1 \leq i \leq n\}$ for every dimension $j$ $(1 \leq j \leq d)$.

    Finally, we notice the connection between the K-center problems (with identical centers of a fixed shape $F$ and fixed sizes) and the problem of computing the largest factor by which we can scale an object of a fixed shape $F$ (e.g. sphere, hyper-rectangle, polyhedron) and initial sizes, such that we can place it inside a bounded domain containing $n$ points and none of the points are located inside the object. The shape $F$ must have a "center" point, around which the scaling is performed. In both problems we can binary search the result $R$ (the minimum maximum distance to the closest center for the K-center problem, or the largest scaling factor for the second problem). Then, we need to perform a feasibility test. For the K-center problem, we need to place an object of shape $F$, centered at each of the $n$ points. Then, we inflate (scale) each object centered at a point $i$ such that it contains all the points at distance at most $R/w(i)$ from it ($w(i)$ is the weight of point $i$). The feasibility test consists of checking if the $n$ inflated shapes are K-pierceable, i.e. if there is a set of $K$ points, such that each of the $n$ inflated shapes contains at least one point from the set. For the second problem, we place an object of the same shape $F$, scaled by the factor $R$, centered at each of the $n$ points (in this case, the points are unweighted). Then, we shrink the domain, such that all the points which cannot be the center of an empty object (scaled by the factor $R$) because some parts of the object would be located outside of the domain are left outside the shrunken domain. Then, we need to check if the $n$ scaled shapes cover the entire shrunken domain. If they don't, then this means that there is at least one point inside the initial domain where we can place the center of an empty object, scaled by the factor $R$, such that it contains no points inside of it (i.e. $R$ is feasible). If the whole domain (area, volume, etc.) is covered by the $n$ scaled objects, then $R$ is not feasible. If $R$ is feasible, we will search for a smaller (larger) value next in the case of the K-center (largest empty scaled object) problem; otherwise, we will search for a larger (smaller) value next.

## 7. RELATED WORK

In [14], the authors discuss a register allocation problem which can be interpreted as a very simple version of the directed tree workflow scheduling problem studied in this paper. Generalizations of the algorithm presented in [14] were given in [6]. Debt management is an important topic nowadays, especially in the context of the global financial crisis. [13] considers the problem of debt relief and repayment capacity of U. S. households from an algorithmic perspective. An optimal resource allocation algorithm which also considers the debt structure of a company is presented in [7]. Conway's Game of Life [9] made cellular automata very popular, which later proved to be important tools for modeling many dynamical, parallel and biological systems. Brief versions of some of the algorithmic techniques presented in this paper (e.g. those for the cellular automaton or for the *Resource Collector 1* game) have previously been presented in a conference short paper [4]. Resource allocation and management problems related to geometric K-center problems were presented in [11]. Efficient algorithms for geometric optimization problems similar to those introduced in this paper were given in [5]. In the future, we intend to consider more complex risk models or cost calculation methods, like those presented in [1] or [12].

## 8. CONCLUSIONS AND FUTURE WORK

In this paper we considered several constrained resource allocation problems, under low risk circumstances. For each problem we developed novel and efficient algorithmic solutions for computing optimal resource allocation strategies. The constraints were formulated as geometric or activity dependency restrictions, or were based on game-theoretic models. As future work, we intend to introduce explicit probability distributions for the values of the parameters of the considered problems, and to not restrict our attention to expected values only.

## REFERENCES

[1] Andreica, M. E., Dobre, I., Andreica, M., Nitu, B., and Andreica, R., "A New Approach of the Risk Project from Managerial Perspective", *Economic Computation and Economic Cybernetics Studies and Research*, vol. 42, 2008, pp. 121-129.
[2] Andreica, M. I., "Efficient Algorithmic Techniques for Several Multidimensional Geometric Data Management and Analysis Problems", *Informatica Economica Journal*, vol. 12, no. 4 (48), 2008, pp. 17-21.
[3] Andreica, M. I., and Tapus, N., "Efficient Data Structures for Online QoS-Constrained Data Transfer Scheduling", *Proceedings of the 7th IEEE International Symposium on Parallel and Distributed Computing,* 2008, pp. 285-292.
[4] Andreica, M. I., and Tapus, N., "Intelligent Strategies for Several Zero-, One- and Two-Player Games", *Proc. of the 4th IEEE Intl. Conf. on Intelligent Computer Communication and Processing*, 2008, pp. 253-256.
[5] Andreica, M. I., Tirsa, E.-D., Andreica, C. T., Andreica, R., Ungureanu, M. A., "Optimal Geometric Partitions, Covers and K-Centers", *Proc. of the 9th WSEAS Intl. Conf. on Mathematics and Computers in Business and Economics*, 2008, pp. 173-178.
[6] Appel, A., and Supowit, K. J., "Generalizations of the Sethi-Ullman Algorithm for Register Allocation", *Software: Practice and Experience*, vol. 17 (6), 1987, pp. 417-421.
[7] Basso, A., and Peccati, L. A., "Optimal resource allocation with minimum activation levels and fixed costs", European *Journal of Operational Research*, vol. 131 (3), 2001, pp. 536-549.
[8] Demaine, E. D., Demaine, M. L., O'Rourke, J., "PushPush and Push-1 are NP-hard in 2D", *Proc. of the 12th Canadian Conference on Computational Geometry*, 2000, pp. 211-219.
[9] Gardner, M., "Mathematical Games: The Fantastic Combinations of John Conway's New Solitaire Game 'Life'", *Scientific American*, vol. 223, 1970, pp. 120–123.
[10] Kataoka, N., Kuroda, K., Ohkawa, T., Koizumi, H., Siratori, N., „Core Business Workflow Model using Internet", *Proc. of the IEEE Intl. Workshops on Parallel Processing*, 1999, pp. 406-411.
[11] Lee, K.-W., Ko, B.-J., Calo, S., "Adaptive Server Selection for Large Scale Interactive Online Games", *Computer Networks*, vol. 49 (1), 2005, pp. 84-102.
[12] Lepadatu, G. V., "The Financial Administration Accountancy Method and the Cost Calculation Method, based on Orders", *Metalurgia International*, vol. 13 (2), 2008, pp. 29-32.
[13] Manning, R. D., "Responsible Debt Relief: An Algorithmic Assessment of Household Debt Capacity and Repayment Capability", *Research Report, Filene Research Institute*, 2008.
[14] Sethi, R., and Ulman, J. D., "The Generation of Optimal Code for Arithmetic Expressions", *Journal of the ACM*, vol. 17 (4), 1970, pp. 715-728.